\documentclass[conference]{IEEEtran}
\IEEEoverridecommandlockouts
\usepackage{cite}
\usepackage{amsmath,amssymb,amsfonts}
\usepackage{algorithmic}
\usepackage{graphicx}
\usepackage{textcomp}
\usepackage{xcolor}
\usepackage{makecell}
\usepackage{colortbl} 
\usepackage{comment}
\usepackage{flushend}
\usepackage{multirow}

\newcommand*{\cussecOld}[1]{\textbf{#1. }}
\newcommand*{\citedata}[1]{\hspace{-3pt}}

\def\BibTeX{{\rm B\kern-.05em{\sc i\kern-.025em b}\kern-.08em
    T\kern-.1667em\lower.7ex\hbox{E}\kern-.125emX}}

\begin{document}
%\title{Analyzing ICSE Technical Paper Utilization and Associated Artifacts: A Decade-long Perspective}
\title{Decade-long Utilization Patterns of ICSE Technical Papers and Associated Artifacts}

% \author{\IEEEauthorblockN{Annonymous Authors}
% \IEEEauthorblockA{\textit{Depts of ABC} \\
% \textit{Institutions}\\
%  A list of Countries\\
%  emails@email.com}
% }
\author{\IEEEauthorblockN{Sharif Ahmed}
\IEEEauthorblockA{\textit{Dept.  of Computer Science } \\
\textit{Boise State University}\\
Boise, USA \\
sharifahmed@u.boisestate.edu}
\and
\IEEEauthorblockN{Rey Ortiz}
\IEEEauthorblockA{\textit{Dept.  of Computer Science} \\
\textit{California Polytechnic State University}\\
San Luis Obispo, USA \\
dortiz11@calpoly.edu}
\and
\IEEEauthorblockN{Nasir U. Eisty}
\IEEEauthorblockA{\textit{Dept.  of Computer Science} \\
\textit{Boise State University}\\
Boise, USA \\
nasireisty@boisestate.edu}
}
%\settopmatter{printfolios=true}

\maketitle
% \IEEEpeerreviewmaketitle 

\begin{abstract}
\textbf{Context:} 
Annually, ICSE acknowledges a range of papers, a subset of which are paired with research artifacts such as source code, datasets, and supplementary materials, adhering to the Open Science Policy. 
However, no prior systematic inquiry dives into gauging the influence of ICSE papers using artifact attributes.
\textbf{Objective:} 
We explore the mutual impact between artifacts and their associated papers presented at ICSE over ten years.
\textbf{Method:} 
% Our approach encompasses collecting data about usage attributes from both papers and their respective artifacts. 
% We then embark on a statistical assessment of these attributes, targeting discernible differences between the attributes of the papers and their affiliated artifacts. 
% In addition, a focused manual analysis is undertaken on the top five papers across each attribute category.
We collect data on usage attributes from papers and their artifacts, conduct a statistical assessment to identify differences, and analyze the top five papers in each attribute category.
\textbf{Results:} 
% Statistically, a noteworthy disparity exists between paper citations and how frequently their associated artifacts are utilized.
% The statistical analyses consistently reveal no significant difference between paper citations and GitHub stars. 
% However, we observe variations in views and/or downloads of the paper and artifact.
% A deep dive into the top five papers across these attributes reveals that papers prominently featuring in both paper and artifact attributes are most frequently associated with criteria such as paper citations, GitHub stars, forks, watches, and the length of the longest issue. 
% This finding somewhat echoes the insights from our statistical analysis.
There is a significant difference between paper citations and the usage of associated artifacts. 
While statistical analyses show no notable difference between paper citations and GitHub stars, variations exist in views and/or downloads of papers and artifacts. 
%A detailed examination of the top five papers indicates that those featuring prominently in both categories are often associated with criteria like paper citations, GitHub stars, forks, watches, and the length of the longest issue, aligning with our statistical analysis.
\textbf{Conclusion:} 
% This endeavor offers a comprehensive overview of ICSE's accepted papers over the past decade, highlighting the nuanced interplay between research papers and their artifacts. 
% To fortify the evaluation of artifact influence within the software research sphere, we advocate for including salient attributes that might be present in one platform but absent in another.
We provide a thorough overview of ICSE's accepted papers from the last decade, emphasizing the intricate relationship between research papers and their artifacts. 
To enhance the assessment of artifact influence in software research, we recommend considering key attributes that may be present in one platform but not in another.

% \section{ Lay Abstract}
% \label{sec_lay_abstract}
% \input{sec_lay_abstract}
\end{abstract}


\section{ Introduction}
\label{sec_introduction}

The International Conference on Software Engineering (ICSE) stands as the premier conference in the field of software engineering, uniting educators and researchers from around the globe to discuss the latest advancements, trends, and insights. 
Each year, a dedicated committee at ICSE meticulously reviews submissions to the technical track, selecting those papers that showcase original contributions and unpublished findings in the realm of Software Engineering.
In a gesture of recognizing exceptional work, committee members nominate and bestow awards upon notable papers, reviewers, and artifacts on an annual basis. 
One of the highlight awards presented is the ICSE N-10 Most Influential Paper Award. This accolade is given to the author(s) of a paper presented at the ICSE conference a decade prior, which has demonstrated a significant impact on either the theory or practice of software engineering over the span of 10 years since its unveiling. 
The recipients of this prestigious award are determined by a specialized team of experts from ICSE.
Despite such recognition, the conference has yet to introduce an equivalent honor for the ICSE N-10 Influential Artifact, nor has it established a quantitative measure, such as paper citation counts, for evaluating research artifacts.
% ICSE, the leading conference in software engineering, brings together global educators and researchers to discuss cutting-edge advancements. 
% A meticulous committee selects papers for the technical track each year, recognizing outstanding contributions with awards. 
% The ICSE N-10 Most Influential Paper Award honors papers that significantly impact software engineering over a decade.
% However, there is currently no equivalent recognition for the ICSE N-10 Influential Artifact, and there is no established quantitative measure for evaluating research artifacts, like paper citation counts.

Referencing paper citations is a well-accepted quantitative method to gauge the significance or impact of research endeavors within the scientific community~\cite{liem2023treat}.
Additionally, the embracement of Open Science Policy standards by preeminent software engineering conferences has notably propelled the sharing and reutilization of research artifacts, such as source code, datasets, or other pertinent research materials~\cite{Mendez_Graziotin_Wagner_Seibold_2020openscienceinse}. 
The availability of these artifacts enables fellow researchers to integrate them into their projects, which, in turn, often leads to the publication of new research papers. 
This cycle beneficially augments the citation count of the original papers that provided the artifacts, enhancing their recognized value within the academic sphere. 
Heumuller et al.~\cite{heumuller2020publish} also found a positive correlation between papers with artifact-link and paper citations. 

Prior works have mostly contrasted the availability, accessibility, and artifact types with paper citations~\cite{frachtenberg2022research,heumuller2020publish,stefan2022retrospective}.
To know more specifically beyond the binary presence or categorical types of paper artifacts, we look into the types of available artifact storage platforms and their publicly available measures of usage and ask the following research questions. 

\textbf{RQ1: Which platforms do researchers use to share artifacts of ICSE papers, and how do they utilize these artifacts?}

\textbf{RQ2: Does the type of artifact platform influence the values of paper usage attributes?}

On the flip side, the publication of new papers leveraging existing artifacts tends to enhance the visibility and accessibility of these artifacts, as evidenced by increased views, downloads, and likes.
Based on this observation, we postulate that the usage dynamics between paper-artifact and paper on digital domains could be related or identical, leading us to pose our research question,

\textbf{RQ3: How do the usage of papers and artifacts compare?}

Furthermore, we observe a variance in the features provided by different artifact-sharing platforms; for instance, GitHub facilitates star allocations but does not track views/downloads, whereas Zenodo tracks views but does not offer a star feature. 
To pinpoint the artifact attributes strongly associated with paper usage, %we dive into our next research questions.
we split our RQ3 into the following research questions. 
% \textbf{RQ2.1: Does artifact-platform type influence the paper usage?}

\textbf{RQ3.1: What is the correlation between artifact attributes and paper usage attributes within platforms?}

\textbf{RQ3.2: Which artifact usage attributes on a platform effectively approximate attributes of other platforms?}

\textbf{RQ3.3: Which artifact attributes, along with their approximations, exhibit correlation with paper usage attributes across platforms?}

To navigate these posed research queries, we employ suitable variants of two sample independent, two sample paired t-tests, and effect size measures as our statistical yardstick and conduct manual scrutiny on the top five papers from our curated dataset, encompassing paper and artifact usage attributes.

In summary, our investigative endeavor examines ICSE research papers from 2011 to 2021, which have publicly available repository links, and scrutinizes their usage attributes as presented by online publishers and the platforms where the papers and artifacts are shared.
This study presents statistical evidence concerning the usage attributes, exploring their interrelationship across two distinct scholarly outputs: research papers and research artifacts.

\begin{comment}
    
\cussecOld{Paper Organization} The rest of the paper is organized as follows. 
% First, Section~\ref{sec_background} provides a brief background of various usage attributes found in digital platforms for papers and their artifacts. 
First, Section~\ref{sec_methodology} 
% provides a brief background of various usage attributes found in digital platforms for papers and their artifacts.
% It also 
describes data types (various usage attributes found in digital platforms for papers and their artifacts), our data collection, data approximation, data derivation process, and analysis setup in response to our research questions.
Second, Section~\ref{sec_result_and_dicussion} reports the demographic information and the findings of our statistical and manual analysis, followed by our discussions.
Third, Section~\ref{sec_related} reviews and contrasts the studies related to our work. 
Finally, Section~\ref{sec_threats} recognizes the threats to the validity of this work, and  Section~\ref{sec_conclusion} concludes it.

\end{comment}

% \section{Background}
% \label{sec_background}
% \input{sec_background}

\section{ Methodology}
\label{sec_methodology}
% In this section, we describe the process of data collection and the methodology for examining statistical evidence.
This section describes the methodology %employed 
for data collection and the approach adopted to analyze the statistical evidence.
\subsection{Data Types}
\label{sec_background}
%In this section, we discuss various usage attributes for research papers and their artifacts shared and consumed on online platforms. 
This subsection presents the diverse usage attributes associated with research papers and their accompanying artifacts as they appear and circulate on digital platforms.

\subsubsection{Paper Usage Attributes}
% We find views, downloads, and citations of the ICSE papers as usage attributes on IEEE Explorer and ACM Digital Library. 

% \cussecOld{Paper Views/Downloads} It shows how many times a paper is accessed or saved by a reader from digital libraries. 

% \cussecOld{Paper Citations} It shows the number of times a paper is cited in other scholarly articles. 
We find views, downloads, and citations of ICSE papers as usage attributes on IEEE Explorer and ACM Digital Library.

\cussecOld{Paper Views/Downloads} This attribute indicates the frequency with which a paper is accessed or downloaded by readers from digital libraries.

\cussecOld{Paper Citations} This attribute reflects the number of times a paper is referenced in other academic works.

\subsubsection{Artifact Usage Attributes}
Artifacts often find their place on platforms such as Zenodo, GitHub, FigShare, or individual websites. 
Notably, well-known artifact-sharing platforms, Zenodo and FigShare, display the count of views and downloads for a given research artifact tied to a research paper. 
In contrast, GitHub, renowned for its extensive
code and data-sharing features, boasts various attributes like issues, commits, stars, forks, and more.

\subsection{Data Collection}
\label{sec_data_collection}

% To examine the influence of ICSE paper citations on artifact usages and vice versa, we conduct an empirical study on available artifacts of ICSE papers from the years 2011 through 2021. 
% We choose to look at artifacts from over a decade as this will give us a good sample size range of data for our study. 
To probe the impact of ICSE paper citations on artifact usage and reciprocally, we carry out an empirical analysis of the accessible artifacts of ICSE papers ranging from the years 2011 to 2021. 
We opt to examine artifacts spanning over a decade as this duration provides a substantial data sample size for our investigation.

% Please RQ1d the following required packages to your document preamble:
% \usepackage{graphicx}
\begin{table}[hbtp]
\caption{Data Extraction Form}
\renewcommand{\arraystretch}{1.3}

\label{tbl_form}
\resizebox{\columnwidth}{!}{%
\begin{tabular}{|p{0.35\linewidth}|p{0.5\linewidth}|p{0.13\linewidth}|}
\hline
\textbf{Field}             & \textbf{Field/Attribute Description}                    & \textbf{Remarks} \\ \hline
Year                       & Number                                                  & RQ1,2,3               \\ \hline
Artifact Name              & Free Text                                               & RQ1                   \\ \hline
Authors                    & Free Text                                               & -               \\ \hline
Title                      & Free Text                                               & -                   \\ \hline
Collected From             & IEEE/ ACM                                               & RQ1                   \\ \hline
Artifact Obtained From     & Paper/ Google/ Zenodo/  Broken                    & RQ1,2                   \\ \hline
Artifact Link              & URL                                                     & RQ1                   \\ \hline
Artifact Link2             & URL                                                     & RQ1                   \\ \hline
Programming-Language  & GitHub Attribute                                        & RQ1                  \\ \hline
\#Issues Open              & GitHub Attribute                                        & RQ3                  \\ \hline
\#Issues Closed            & GitHub Attribute                                        & RQ3                  \\ \hline
Longest Issue              & GitHub Attribute                                        & 
RQ3                  \\ \hline
\#Pull Request Open              & GitHub Attribute                                        & RQ3                  \\ \hline
\#Pull Request Closed            & GitHub Attribute                                        & RQ3                  \\ \hline
Longest Pull Request              & GitHub Attribute                                        & 
RQ3                  \\ \hline
\#Contributors                  & GitHub Attribute                                        & RQ3   \\ \hline
\#Commits                  & GitHub Attribute                                        & RQ3                  \\ \hline
Watching                   & GitHub Attribute                                        & RQ3                  \\ \hline
Forks                      & GitHub Attribute                                        & RQ3                  \\ \hline
Last Commit Date           & GitHub Last Activity                                    & RQ3                  \\ \hline
Conference Diff Date       & \# Days  Between Last Commit Date and Conference Date   & RQ3                  \\ \hline
Stars                      & GitHub Attribute                                        & RQ3                  \\ \hline
Paper Views/DownloRQ1s      & IEEE/ACM Attribute                                      & RQ2,3                  \\ \hline
Paper Citations            & IEEE/ACM Attribute                                      & RQ2,3                  \\ \hline
Conference Date            & Venue Start Date                                        & RQ3                  \\ \hline
Artifact DownloRQ1s         & Zenodo \& FigShare Attribute                            & RQ3                  \\ \hline
Artifact Views             & Zenodo \& FigShare Attribute                            & RQ3                  \\ \hline
Has Discussion             & Derived                                                 & RQ3                  \\ \hline
Overall Artifact Usage     & Derived                                                 & RQ3             \\ \hline
Normalized Artifact Usage  & Normalized Artifact Usage wrt. creation year \& 2023    & RQ3             \\ \hline
Normalized Paper Citations & Normalized Paper Citation wrt. publication year \& 2023 & RQ2,3             \\ \hline

\end{tabular}%
} %
% \footnotesize{ Here, P/A-D: Paper/Artifact-Demographics,  RQ: Research Question }
\end{table}
\subsubsection{Paper Selection}
\label{sec_paper_selection}
% ICSE's official website (\textcolor{blue}{https://conf.researchr.org/series/icse}) shows the list of accepted technical full papers with artifact-evaluation badges. 
% Research papers that have been accepted receive a "badge" that certifies a paper is from the following forms: Functional, Reusable, Available, Replicated, and Reproduced. 
% Accepted research papers that have the green "Available" badge, have been identified to have a publicly accessible repository through a DOI or link. 
%  A DOI or link to the repository is provided in the research paper. 
% % The types of repository links can be from the following but not limited to Github, Zenodo, Figshare, or Google Code Archive all of which include diverse attributes.
% 
The official ICSE website\footnote{https://conf.researchr.org/series/icse} displays a list of accepted technical full papers, accompanied by artifact-evaluation badges. 
These badges are conferred upon accepted research papers, indicating their status in one of the following categories: Functional, Reusable, Available, Replicated, or Reproduced. 
Among these, papers adorned with the green ``Available" badge are recognized to possess publicly accessible repositories, accessible via a DOI or link.

% Particularly, we include artifact `Available' papers and exclude the ``Artifact-Evaluation'' track as they have papers from different years.
% So, we select the papers from the official website from the year 2021 down to 2011.
% For ICSE-2018, we find the official website link broken; then, we turn to the ACM and IEEE websites for ICSE proceedings for 2018 through 2011.
% Unfortunately, we find these badges are unavailable for the year 2017 and before. 
% Thus, we manually explore the accepted technical papers and look for the artifact links inside the papers.
% Since there is only one paper that has a badge from the year 2018, we decide to manually explore the papers accepted in 2018 as well.
% If a paper contains an artifact link, we select that paper. 
% Otherwise, we search for the artifact on Google, currently the largest search engine,  with the paper title. 
% Additionally, we try to find the paper-artifact on Zenodo,
% the artifact-sharing platform that is popular and has a view/download attribute. 
% We list the papers that are marked with badges or are selected by our manual lookup.
Mainly, we focus on papers with the `Available' artifact badge, while we omit those from the ``Artifact-Evaluation" track due to including papers from varied years. 
Thus, we sift through papers on the official website, starting from 2021 and tracing back to 2011.
However, we encountered an issue with ICSE-2018: the official website link was non-functional. 
We consulted the ACM and IEEE websites to access ICSE proceedings from 2018 to 2011 as a workaround. 
To our disappointment, these badges were absent for years before 2018. 
Consequently, we manually reviewed the accepted technical papers, looking for embedded artifact links. 
Given that only a single paper from 2018 bore the badge, we also undertook a detailed manual examination of the 2018 accepted papers.

Our criteria for paper selection revolved around the presence of an artifact link. 
In its absence, our next step was to leverage Google, the predominant search engine, by entering the paper title to locate the associated artifact. 
Furthermore, we tried to locate the paper-artifact on Zenodo, a renowned artifact-sharing platform that conveniently offers view/download statistics. 
Papers either carrying the designated badges or identified through our 
% hands-on 
manual 
search have been enumerated in our study.

\subsubsection{Data Extraction}
% After selecting the papers, we plan to extract the paper attributes from the paper publishers such as \textit{paper-citation }and\textit{ paper-views}.
% Next, we find the artifacts are published on different platforms which provide different artifact consumption attributes. 
% To analyze these attributes, we develop a data extraction form as described in Table ~\ref{tbl_form}
%  following a work~\cite{ahmed2022automatic}. 
% Here, we collect the \textit{title, year, author,  published on ACM/IEEE, conference date, views, downloads, }and\textit{ citations} of the selected papers. 
% For artifacts of the selected papers, we collect 
% \textit{number of issues open/closed, longest issue }(the issue with maximum comments and replies)\textit{, number of commits,  last commit date }(last activity), and  if they are shared on GitHub\footnote{https://github.com} or Google Code Archive\footnote{https://code.google.com/archive}.
% We also collect \textit{stars, forks, watching, }
% and \textit{programming language used} 
% for artifacts shared on GitHub. 
% For artifacts of selected papers shared on Zenodo\footnote{https://zenodo.org} or Figshare\footnote{https://figshare.com}, 
% we collect \textit{views }and\textit{ downloads} of the artifact.
After identifying the appropriate papers, our next step is to extract relevant attributes from the publishers, focusing on aspects such as \textit{paper-citation} and \textit{paper-views}. 
We note that artifacts are often disseminated across diverse platforms, each offering diverse attributes related to artifact engagement.
To systematically evaluate these attributes, we craft a data extraction form, detailed in Table ~\ref{tbl_form}, taking inspiration from a previous study~\cite{ahmed2022automatic}. 
Specifically, for each paper, we collate data on its \textit{title, year, author, publication source (ACM/IEEE), conference date, views, downloads}, and \textit{citations}.
When considering the artifacts associated with these papers, we gather data on the \textit{number of open/closed issues, the most extensively discussed issue} (characterized by the maximum number of comments and replies), \textit{number of commits, date of the last commit} (indicating recent activity) if these artifacts are hosted on platforms like GitHub\footnote{https://github.com} or Google Code Archive\footnote{https://code.google.com/archive}.
Precisely, we extract additional attributes for artifacts hosted on GitHub, including \textit{stars, forks, watching}, and the \textit{programming language used}. 
Conversely, for those hosted on Zenodo\footnote{https://zenodo.org} or Figshare\footnote{https://figshare.com}, our focus shifts to collecting data on the \textit{views} and \textit{downloads} of the artifact.

\subsection{Data Processing}
\label{sec_data_processing}
Here, we describe how we approximate the missing data, derive new data, and normalize yearly accumulated data of attributes from paper and artifact usages. 

\subsubsection{Data Approximation}
\label{sec_artifact_approximation}
% Since the artifact-sharing platforms provide different attributes and there are attribute(s) available in one platform but unavailable in another, we plan to approximate the missing attribute values from paper that are shared on multiple platforms.
% To obtain approximate attribute values, we adopt the simple \textit{Unitary method}. 
% For example, if there is an artifact with a total of 85\textit{ artifact views} and 10 GitHub \textit{stars}, we first divide 85 by 10 and get an 8.5 artifact \textit{views~stars} factor. 
% Second, we obtain such \textit{views~stars} factor for each row that has both original \textit{artifact-views} and \textit{stars}.
% Third, we get the median of the \textit{views~stars} factor as a final value to approximate the GitHub stars from artifact views. 
% Fourth, we approximate the views~stars for each row that has views but no values for GitHub stars.
% We follow this approach for every pair of artifact usage attributes addressed in Table~\ref{tbl_form}. 
% Finally, we take the mean of all approximate attribute values and name them `approximate artifact usage'. 
Owing to the varied attributes provided by different artifact-sharing platforms and the presence of specific attributes on one platform but their absence on another, we intend to estimate the missing attribute values for papers shared across multiple platforms. 
We employ the straightforward \textit{Unitary Method} to approximate these missing values. 
For instance, consider an artifact that has accumulated 85 \textit{artifact views} and 10 GitHub \textit{stars}. 
As a starting point, by dividing 85 by 10, we derive an 8.5 \textit{views-per-star} ratio. 
Next, for every entry possessing both the original \textit{artifact-views} and \textit{stars}, we determine a similar \textit{views-per-star} ratio.
We calculate the median from these ratios, which serves as our established conversion rate to predict GitHub stars from artifact views.
Subsequently, we apply this conversion rate to every entry with available views data lacking star counts.
% This methodology is consistently employed for all corresponding pairs of artifact usage attributes outlined in Table~\ref{tbl_form}.
This methodology is consistently employed for all corresponding pairs of artifact usage attributes from GitHub and non-GitHub outlined in Table~\ref{tbl_form}.
However, such approximation is not applied to any paper usage attributes for their uniform availability.
% Next, we select the most associated approximated attribute values using Mann Whitney U test~\cite{mann1947test} and Cliff's Delta ($\delta$)~\cite{cliff1993dominance}.
% Ultimately, we compute the average of all estimated attribute values, designating this aggregate as the `approximate artifact usage.'

\subsubsection{Data Derivation}
% After extracting and approximating the usage attribute values, we derive additional data that may provide new insights.
% To check the post-publication influence of an artifact, we compute the day differences between the conference start day and the last commit day in the artifact repository. 
% Also, to gauge the engagement of people with or maintenance of the artifacts we check if there are any current or resolved discussions (i.e., issues).
Upon gathering and estimating the usage attribute values, we generate supplemental data that could shed light on deeper insights. 
To assess the enduring influence of an artifact post-publication, we calculate the period between the conference's commencement and the most recent commit date in the artifact repository. 
Moreover, to measure the level of engagement or the upkeep of the artifacts, we examine whether ongoing or previously resolved discussions are represented as issues.

\subsubsection{Data Normalization}
% The paper or artifact attribute values are accrued from different years.
% Older papers are supposed to get higher attribute values than the recent ones.
% So, we choose to normalize them with a division by the length of the time from the publication year to the current year, 2023.
Attribute values for papers or artifacts span across various years. Logically, older papers are anticipated to have accumulated higher attribute values compared to more recent ones. 
Therefore, to level the playing field, we opt to normalize these values by dividing them by the duration from their year of publication up to the %present 
year, 2023.

\subsection{Experiment}
This subsection describes the statistical tools and their configurations, explains the choice of statistical tools for our interrogated research questions, and briefs a complementary manual analysis procedure of our curated data.

\subsubsection{Statistical Measure}
% After having the original, approximate, and normalized paper and artifact attributes in Sec~\ref{sec_data_collection} and Sec~\ref{sec_data_processing}, we perform the statistical two samples paired \textit{t}-test to answer our research questions addressed in Sec~\ref{sec_introduction}.  
% To achieve this, we utilize SciPy's\footnote{https://docs.scipy.org/doc/scipy/reference/generated/scipy.stats.ttest\_rel.html} \textit{t}-test API on two related samples~\cite{2020SciPy-NMeth}.
% Here, we choose \textit{two-sided} alternative hypothesis and omit non-numerical attribute values.
% Next, we run paired \textit{t}-tests by selecting one attribute from paper usage and another attribute from artifact usage following the experiment setup outlined in Sec~\ref{sec_rq}. 
% % for our two research questions.
After collating the original, approximate, and normalized attributes for both papers and artifacts as detailed in Sec~\ref{sec_data_collection} and Sec~\ref{sec_data_processing}, we proceed to apply the paired or independent two-sample \textit{t}-test to address the research questions posited in Sec~\ref{sec_introduction}. 
As the data are not normally distributed, we consider non-parametric versions of both paired and independent two-sample t-tests.
To implement this, we turn to the non-parametric \textit{t}-test functionality provided by SciPy for 
Mann-Whitney U test\footnote{https://docs.scipy.org/doc/scipy/reference/generated/scipy.stats.mannwhitneyu.html}~\cite{mann1947test}
and 
Wilcoxon signed-rank test\footnote{https://docs.scipy.org/doc/scipy/reference/generated/scipy.stats.wilcoxon.html}~\cite{wilcoxon1970critical} to analyze two correlated samples, as referenced in~\cite{2020SciPy-NMeth}. 
For both types of t-tests, we adopt a \textit{two-sided} alternative hypothesis within this context, considering non-numerical attribute values.
We then analyze these usages using the paired or independent non-parametric two-sample t-tests to identify any statistical evidence pointing to differences. 
The robustness of the evidence is assessed via the p-values from the significance test, employing threshold benchmarks of 0.05 (5\%), 0.01 (1\%), and 0.001 (0.1\%).
Moreover, to gauge the strength of the significant differences, we measured their effect-sizes using Cliff's Delta ($\delta$)~\cite{cliff1993dominance} for independent and and Cohen's D~\cite{cohen2013statistical} for paired \textit{t}-tests.
Subsequently, we execute paired \textit{t}-tests by pairing one attribute from paper usage with another from artifact usage and execute independent \textit{t}-tests by taking two unrelated samples from an attribute, adhering to the experimental structure delineated in Sec~\ref{sec_rq}.

\subsubsection{Answering Research Questions}
\label{sec_rq}
Here, we describe our approaches to answer the research questions from Sec~\ref{sec_introduction}.

\textbf{RQ1: Which platforms do researchers use to share artifacts of ICSE papers, and how do they utilize these artifacts?}

To know about the artifacts of ICSE papers from 10 years, we analyze and summarize the demographics of our collected and extracted data from artifact-sharing platforms and digital publishers of papers (described in Sec~\ref{sec_data_collection} and Sec~\ref{sec_data_processing}). 

\textbf{RQ2: Does the type of artifact platform influence the values of paper usage attributes?}

As the artifacts of ICSE papers are stored in different artifact-sharing platforms (e.g., GitHub and Figshare/Zenodo), they get a variety of interactions from different visitors due to the platform's diversity.  
To know the answer to this research question, we check the differences in the paper's impact attributes between two major types of artifact platforms using suitable non-parametric Mann-Whitney U Test~\cite{mann1947test} followed by their effect size measure using Cohen's D~\cite{cohen2013statistical}.
This answer will help us to know if any platform influences the overall impact of the papers. 

\textbf{RQ3.1: What is the correlation between artifact attributes and paper usage attributes within platforms?}

To answer this RQ, we consider paper-views/downloads (aka. full paper views), paper-citations, and normalized paper-citations for paper usage.
For artifact usage, we consider all of the attributes for each of the observed artifact platforms (e.g., GitHub, Zenodo, etc.) individually.
As we consider the artifact and paper usage platform-wise, there are no missing data points.
Next, we select \textit{Wilcoxon Signed-Rank Test}~\cite{wilcoxon1970critical} to test the statistical significance and \textit{Cohen's D} to measure the effect size of the attribute values one from the paper usage and another from the artifact usage.
To complement, we additionally check the Pearson Correlation of these attribute values from two groups.

\textbf{RQ3.2: Which artifact usage attributes on a platform effectively approximate attributes of other platforms?}

To elucidate the findings related to this specific research question, we approximate the attribute values of one type of platform(i.e., GitHub and non-GitHub artifact storage) from another type of platform as described in Sec~\ref{sec_artifact_approximation}.
With the statistical evidence obtained from the answer to the \textbf{RQ3.1}, we approximate only the attributes that are statistically correlated to the paper citations or paper-views/downloads.
Next, we check the statistical differences between the collected original and our approximated values using the Mann-Whitney U test~\cite{mann1947test}.
Then, to capture the strength of the statistical differences, we select Cliff's Delta ($\delta$)~\cite{cliff1993dominance} effect size measure and classify the size following Hess \& Kromrey~\cite{hess2004robust}. 
% The paper usage attributes and artifact usage attributes from non-GitHub platforms are not as many as GitHub. 
% From non-GitHub platforms, we considered both artifact views and artifact download attributes.

\textbf{RQ3.3: Which artifact attributes, along with their approximations, exhibit correlation with paper usage attributes across platforms?}

From knowing the platform-wise artifact attributes in RQ3.1 and their good approximations from other platforms in RQ3.2, we are now poised to gain this research question. 
In the pursuit of knowing which artifact attributes and their approximations are related to paper usage attributes, we select the already evident artifact attributes in the answers to RQ3.1 and RQ3.2.
Next, we take their original and approximated missing values and their normalized values (as we normalized paper citations in Sec~\ref{sec_data_processing}3). 
Finally, we implement a statistical analysis akin to the approach for RQ3.1, juxtaposing the values of each artifact attribute with the attributes of paper usage. 
Here, we are taking the total samples from paper and artifact attributes instead of sampling platform-wise as RQ3.1.

The answers to the RQ3.1, RQ3.2, and RQ3.3 collectively answers their parent research questions, \textit{RQ3: How do the usage of papers and artifacts compare?} 

\subsubsection{Manual Analysis}
\label{meth_manual_analysis}
Alongside these statistical approaches to answer the research questions (Sec~\ref{sec_rq}), we perform a hands-on analysis, selecting the top five papers ( equivalent to top 2\% of our datasets) for each artifact attribute as outlined in Table~\ref{tbl_form}.
Here, we consider the attribute-wise appearances of paper or artifacts.
Next, we see how these papers or their artifacts co-appear in the top lists.
Then, we scrutinize the top data points from each attribute derived from all observed platforms.
Lastly, we try to infer our manual analysis and contrast it with pieces of evidence collected from the aforementioned statistical measures.

\section{ Results and Discussion}
\label{sec_result_and_dicussion}
This section reports and discusses the demographics of our collected data and the results of our experiments.
\begin{figure}[htbp]
    \centerline{\includegraphics[trim={0 0cm 0 0.0cm},clip, width=.85\linewidth]{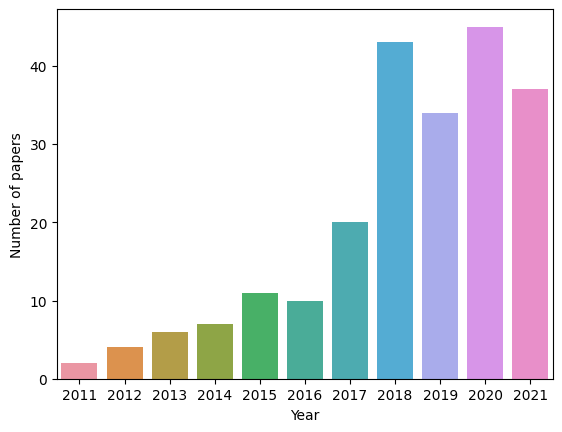}}

    \caption{Artifacts found in Papers Accepted at ICSE 
    from 2011 to 2021
    }
    \label{fig_paper_with_artifacts}
\end{figure}
% \begin{figure}[hbtp]
%     \centerline{\includegraphics[trim={0 0cm 0 0.0cm},clip, width=.85\linewidth]{res/fig/demo-both-V.pdf}}

%     \caption{ Paper Usage and Artifact Usage of ICSE Papers
%     % from 2011 to 2021
%     }
%     \label{fig_usage_both}
% \end{figure}
% \begin{figure*}[hbtp]
%     \centerline{\includegraphics[trim={0 0cm 0 0.0cm},clip, width=1\linewidth]{res/fig/demo-both.png}}

%     \caption{ Paper Usage and Artifact Usage of ICSE Papers
%     % from 2011 to 2021
%     }
%     \label{fig_usage_both}
% \end{figure*}

% \begin{figure}[hbtp]
% \centering
    
\begin{figure}%{\linewidth}
    \centering
    \includegraphics[width=.85\linewidth]{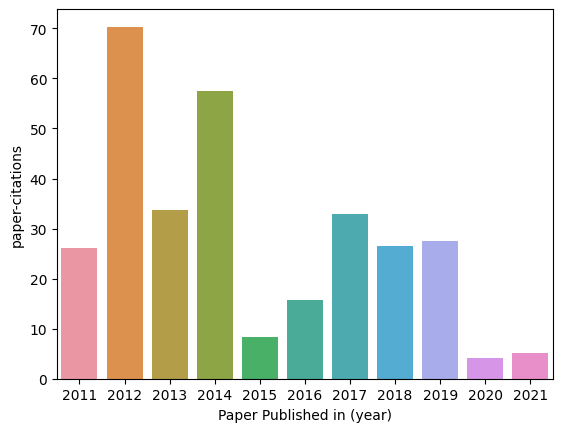}
    \caption{Citations of ICSE Artifact-Papers from 2011 to 2021}
    \label{fig-a}
\end{figure}
% \begin{figure}%{\linewidth}
%     \centering
%     \includegraphics[width=.85\linewidth]{res/fig/bars/bar_normalized_paper.png}
% \end{figure}
% \begin{figure}%{\linewidth}
%     \centering
%     \includegraphics[width=.85\linewidth]{res/fig/bars/bar_artifact_overall.png}
% \end{figure}
% \begin{figure}%{\linewidth}
%     \centering
%     \includegraphics[width=.85\linewidth]{res/fig/bars/bar_artifact_overall_normalized.png}
% \end{figure}
\begin{figure}%{\linewidth}
    \centering
    \includegraphics[width=\linewidth]{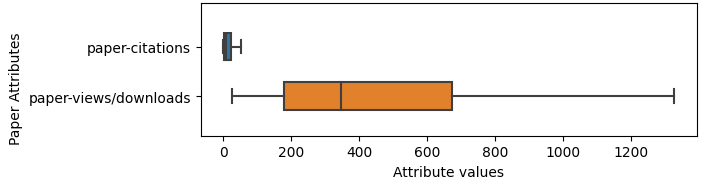}
    \caption{Paper Usage of ICSE Papers}
    \label{fig-b}
\end{figure}
\begin{figure}%{\linewidth}
    \centering
    \includegraphics[width=\linewidth]{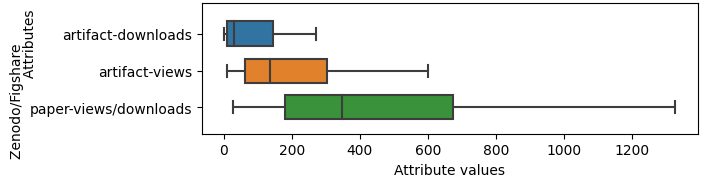}
    \caption{Artifact Usage (Zenodo/ FigShare)}
    \label{fig-c}
\end{figure}
\begin{figure}%{\linewidth}
    \centering
    \includegraphics[width=\linewidth]{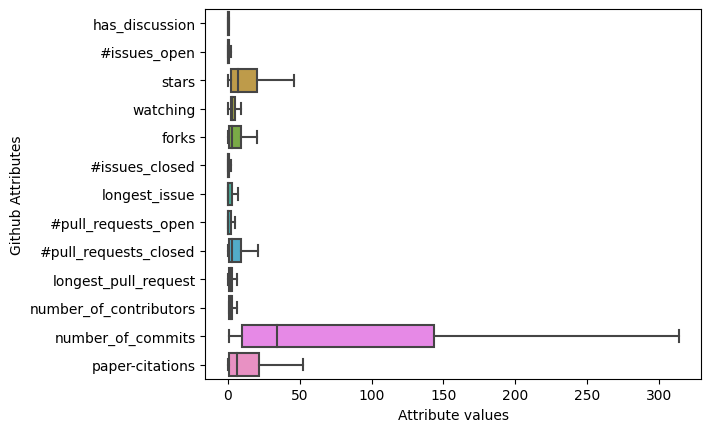}
    \caption{Artifact Usage (GitHub)}
    \label{fig-d}
\end{figure}
%     \caption{ Paper Usage and Artifact Usage of ICSE Papers
%     % from 2011 to 2021
%     }
%     \label{fig_usage_both}
% \end{figure}
% \textbf{Demographic Information}
\textbf{RQ1: Which platforms do researchers use to share artifacts of ICSE papers, and how do they utilize these artifacts?}

% In our data collection process, we find 234 papers that have artifacts. 
% 198 of these papers contain links to the artifacts. 
% We find these links: Broken (8), GitHub (170), Zenodo(18), FigShare (4), Google-Code (5), and no-artifact (1).
% Fig.~\ref{fig_paper_with_artifacts} shows the number of papers accepted at ICSE with artifacts has doubly increased after 2017, the year before artifact badges became available. 
% We see these numbers are higher for even years than odd years after 2017.
% Notably, the papers accepted in even years are published in ACM Digital Library and the papers accepted in odd years are published in IEEE Explorer.
% The top 10 used programming languages in GitHub artifacts are Java, Python, JavaScript, C, C++, R, OCamel, Scala, Shell, and Go.
During our data collection, we identified 234 papers that included artifacts. 
Out of these, 198 papers provide direct links to their associated artifacts. 
The breakdown of these links is as follows: 8 were broken, 170 led to GitHub, 18 to Zenodo, 4 to FigShare, 5 to Google Code, and 1 link did not lead to any artifact. 
As depicted in Fig.~\ref{fig_paper_with_artifacts}, there is a discernible increase in the number of ICSE-accepted papers featuring artifacts post-2017, the year preceding the %introduction of artifact badges. 
first noticed `Available' badge on an ICSE paper.
Interestingly, post-2017, even-numbered years exhibit higher counts than odd-numbered years. 
It is worth noting that papers accepted during even years get published in the ACM Digital Library, while those from odd years are featured in IEEE Explorer.
However, we do not find such a pattern for the citation of papers in Fig.~\ref{fig-a}. 
Also, Fig.~\ref{fig-b} shows that paper views/ downloads (aka. full paper views) are greatly higher and more variable than paper citations.
Fig.~\ref{fig-c} shows similar differences and variability for artifact views and downloads, but these are lower than paper views/ downloads.

A possible reason is that the paper is viewed, skimmed, and read for many reasons, whereas it is cited if its contribution is used in another paper. 
Similarly, an artifact is downloaded if it is of any potential interest to the viewer of the artifact.
So, artifact views are higher than their downloads.
Conversely, GitHub does not capture view/ download counts but a wide span of other attributes that have a much lower range of values (see Fig.~\ref{fig-d}).
Because these attributes are not auto-enumerated as views or download actions and require additional interaction from the owner or visitor of the artifact.
When examining the programming languages utilized in GitHub artifacts, the top 10 in descending order are Java, Python, JavaScript, C, C++, R, OCamel, Scala, Shell, and Go.

\begin{table*}[hbtp]
\caption{ Paper vs Artifact Usages within Artifact Platform}
\label{tab:within_platform}
\renewcommand{\arraystretch}{1.25}
\resizebox{\linewidth}{!}{
\begin{tabular}{p{0.05\linewidth}
 p{0.2\linewidth}p{0.08\linewidth} 
p{0.08\linewidth} p{0.08\linewidth}p{0.08\linewidth} 
p{0.08\linewidth} p{0.08\linewidth} p{0.08\linewidth} 
p{0.08\linewidth} p{0.08\linewidth} 
}
 &
   &
  \multicolumn{3}{c}{\textbf{p-value(Wilcoxon Signed-Rank Test)}} &
  \multicolumn{3}{c}{\textbf{effect-size(Cohen's D)}} &
  \multicolumn{3}{c}{\textbf{Pearson Correlation}} \\
 &
  \multicolumn{1}{r}{\textbf{Paper attributes$\rightarrow$}} &
  \multicolumn{1}{p{0.05\linewidth}}{{$\hat{citation}$}} &
  \multicolumn{1}{p{0.05\linewidth}}{{citation}} &
  \multicolumn{1}{p{0.05\linewidth}}{{full views}} &
  \multicolumn{1}{p{0.05\linewidth}}{{$\hat{citation}$}} &
  \multicolumn{1}{p{0.05\linewidth}}{{citation}} &
  \multicolumn{1}{p{0.05\linewidth}}{{full views}} &
  \multicolumn{1}{p{0.05\linewidth}}{{$\hat{citation}$}} &
  \multicolumn{1}{p{0.05\linewidth}}{{citation}} &
  \multicolumn{1}{p{0.05\linewidth}}{{full-views}} \\
\multicolumn{1}{p{0.0\linewidth}}{\textbf{Platform}} &
  \multicolumn{1}{l}{ ~\textbf{Artifact attributes $\downarrow$}} &
   &
   &
   &
  \multicolumn{1}{l}{} &
  \multicolumn{1}{l}{} &
  \multicolumn{1}{l}{} &
  \multicolumn{1}{l}{} &
  \multicolumn{1}{l}{} &
  \multicolumn{1}{l}{} \\ \hline
GitHub &
  \#issues\_closed &
  \cellcolor[HTML]{FFFFFF}$\ast \ast$ &
  \cellcolor[HTML]{FFFFFF}$\ast \ast \ast$ &
  $\ast \ast \ast$ &
  \cellcolor[HTML]{DAF0E6}0.141568 &
  \cellcolor[HTML]{FCF0EF}-0.362092 &
  \cellcolor[HTML]{E67E75}-1.20434 &
  \cellcolor[HTML]{F7FCFA}0.103702 &
  \cellcolor[HTML]{F7FCF9}0.10671 &
  \cellcolor[HTML]{FEFDFD}0.046882 \\
 &
  \#issues\_open &
  \cellcolor[HTML]{FFFFFF}$\ast \ast \ast$ &
  \cellcolor[HTML]{FFFFFF}$\ast \ast \ast$ &
  $\ast \ast \ast$ &
  \cellcolor[HTML]{F1FAF6}-0.106544 &
  \cellcolor[HTML]{F5CCC8}-0.63192 &
  \cellcolor[HTML]{E67C73}-1.213699 &
  \cellcolor[HTML]{F8FDFA}0.099351 &
  \cellcolor[HTML]{F9FDFB}0.092618 &
  \cellcolor[HTML]{FCFEFD}0.075578 \\
 &
  \#pull\_requests\_closed &
  \cellcolor[HTML]{FFFFFF}- &
  \cellcolor[HTML]{FFFFFF}$\ast \ast \ast$ &
  $\ast \ast \ast$ &
  \cellcolor[HTML]{CBEADB}0.309048 &
  \cellcolor[HTML]{EBF7F1}-0.040344 &
  \cellcolor[HTML]{EA958E}-1.032508 &
  \cellcolor[HTML]{FBEDEB}-0.087446 &
  \cellcolor[HTML]{FBECEB}-0.088204 &
  \cellcolor[HTML]{FBEAE9}-0.106818 \\
 &
  \#pull\_requests\_open &
  \cellcolor[HTML]{FFFFFF}$\ast \ast \ast$ &
  \cellcolor[HTML]{FFFFFF}$\ast \ast \ast$ &
  $\ast \ast \ast$ &
  \cellcolor[HTML]{F5CECA}-0.616239 &
  \cellcolor[HTML]{EDA49E}-0.921565 &
  \cellcolor[HTML]{E98F88}-1.075262 &
  \cellcolor[HTML]{FCF3F2}-0.036361 &
  \cellcolor[HTML]{FCF3F2}-0.037212 &
  \cellcolor[HTML]{FCF1F0}-0.050672 \\
 &
  forks &
  \cellcolor[HTML]{FFFFFF}$\ast \ast \ast$ &
  \cellcolor[HTML]{FFFFFF}$\ast \ast \ast$ &
  $\ast \ast \ast$ &
  \cellcolor[HTML]{D6EFE3}0.185026 &
  \cellcolor[HTML]{E3F4EC}0.046146 &
  \cellcolor[HTML]{E88980}-1.124514 &
  \cellcolor[HTML]{FEFAFA}0.021336 &
  \cellcolor[HTML]{FEFBFB}0.031129 &
  \cellcolor[HTML]{FEFDFD}0.04409 \\
 &
  has\_discussion &
  \cellcolor[HTML]{FFFFFF}$\ast \ast \ast$ &
  \cellcolor[HTML]{FFFFFF}$\ast \ast \ast$ &
  $\ast \ast \ast$ &
  \cellcolor[HTML]{F3C4C0}-0.689841 &
  \cellcolor[HTML]{F2BAB6}-0.757438 &
  \cellcolor[HTML]{E67C73}-1.22024 &
  \cellcolor[HTML]{E6F5EE}0.20265 &
  \cellcolor[HTML]{E5F5ED}0.204936 &
  \cellcolor[HTML]{ECF7F2}0.169922 \\
 &
  longest\_issue &
  \cellcolor[HTML]{FFFFFF}$\ast \ast \ast$ &
  \cellcolor[HTML]{FFFFFF}$\ast \ast \ast$ &
  $\ast \ast \ast$ &
  \cellcolor[HTML]{F8FCFA}-0.175443 &
  \cellcolor[HTML]{F4C9C5}-0.653075 &
  \cellcolor[HTML]{E67C73}-1.214729 &
  \cellcolor[HTML]{FEFFFE}0.068911 &
  \cellcolor[HTML]{FFFFFF}0.061709 &
  \cellcolor[HTML]{FEFEFE}0.053928 \\
 &
  longest\_pull\_request &
  \cellcolor[HTML]{FFFFFF}- &
  \cellcolor[HTML]{FFFFFF}$\ast \ast \ast$ &
  $\ast \ast \ast$ &
  \cellcolor[HTML]{DAF0E6}0.142977 &
  \cellcolor[HTML]{F6D0CD}-0.598908 &
  \cellcolor[HTML]{EA9189}-1.064087 &
  \cellcolor[HTML]{FAE9E8}-0.112594 &
  \cellcolor[HTML]{FAE7E5}-0.13435 &
  \cellcolor[HTML]{FBEAE9}-0.10627 \\
 &
  number\_of\_commits &
  \cellcolor[HTML]{FFFFFF}$\ast \ast \ast$ &
  \cellcolor[HTML]{FFFFFF}$\ast \ast \ast$ &
  $\ast \ast \ast$ &
  \cellcolor[HTML]{D4EEE1}0.210263 &
  \cellcolor[HTML]{D5EEE2}0.202768 &
  \cellcolor[HTML]{EBF7F1}-0.03807 &
  \cellcolor[HTML]{FCF4F3}-0.02877 &
  \cellcolor[HTML]{FDF7F6}-0.004808 &
  \cellcolor[HTML]{FDF8F8}0.006214 \\
 &
  number\_of\_contributors &
  \cellcolor[HTML]{FFFFFF}- &
  \cellcolor[HTML]{FFFFFF}$\ast \ast \ast$ &
  $\ast \ast \ast$ &
  \cellcolor[HTML]{FAFDFC}-0.198792 &
  \cellcolor[HTML]{F4C8C4}-0.656675 &
  \cellcolor[HTML]{E67F77}-1.19126 &
  \cellcolor[HTML]{FCF3F3}-0.031804 &
  \cellcolor[HTML]{FDF5F4}-0.019006 &
  \cellcolor[HTML]{FDF5F5}-0.01664 \\
 &
  stars &
  \cellcolor[HTML]{FFFFFF}$\ast \ast \ast$ &
  \cellcolor[HTML]{FFFFFF}- &
  $\ast \ast \ast$ &
  \cellcolor[HTML]{B4E1CB}0.553973 &
  \cellcolor[HTML]{DCF1E7}0.1226 &
  \cellcolor[HTML]{E7847C}-1.157918 &
  \cellcolor[HTML]{BEE5D2}0.425383 &
  \cellcolor[HTML]{C9EADA}0.360838 &
  \cellcolor[HTML]{D1EDDF}0.317138 \\
 &
  watching &
  \cellcolor[HTML]{FFFFFF}$\ast \ast$ &
  \cellcolor[HTML]{FFFFFF}$\ast \ast \ast$ &
  $\ast \ast \ast$ &
  \cellcolor[HTML]{DEF2E8}0.098318 &
  \cellcolor[HTML]{F6D2CE}-0.587547 &
  \cellcolor[HTML]{E67D74}-1.210433 &
  \cellcolor[HTML]{D9F0E5}0.273896 &
  \cellcolor[HTML]{E1F3EA}0.23086 &
  \cellcolor[HTML]{FEFEFE}0.050604 \\ \hline
Zenodo &
  artifact-downloads &
  $\ast \ast \ast$ &
  $\ast \ast \ast$ &
  $\ast \ast \ast$ &
  \cellcolor[HTML]{C6E8D8}0.356923 &
  \cellcolor[HTML]{C7E9D8}0.344337 &
  \cellcolor[HTML]{FDF4F4}-0.331153 &
  \cellcolor[HTML]{FBEEED}-0.077392 &
  \cellcolor[HTML]{FBEDEC}-0.084037 &
  \cellcolor[HTML]{FDF6F6}-0.009423 \\
 &
  artifact-views &
  $\ast \ast \ast$ &
  $\ast \ast \ast$ &
  $\ast \ast \ast$ &
  \cellcolor[HTML]{B2E0C9}0.57737 &
  \cellcolor[HTML]{B4E1CB}0.555872 &
  \cellcolor[HTML]{F8DCDA}-0.507905 &
  \cellcolor[HTML]{FAFDFC}0.088881 &
  \cellcolor[HTML]{F9FDFB}0.093003 &
  \cellcolor[HTML]{F0F9F5}0.143155 \\ \hline
Figshare &
  artifact-downloads &
  $\ast$ &
  $\ast$ &
  - &
  \cellcolor[HTML]{7ECBA6}1.130097 &
  \cellcolor[HTML]{81CCA8}1.098378 &
  \cellcolor[HTML]{FDF7F6}-0.315432 &
  \cellcolor[HTML]{F1FAF6}0.137243 &
  \cellcolor[HTML]{B7E2CD}0.46376 &
  \cellcolor[HTML]{C0E6D3}0.413977 \\
 &
  artifact-views &
  $\ast$ &
  $\ast$ &
  - &
  \cellcolor[HTML]{57BB8A}1.546925 &
  \cellcolor[HTML]{59BC8B}1.531921 &
  \cellcolor[HTML]{9AD6B9}0.831221 &
  \cellcolor[HTML]{C9EADA}0.362175 &
  \cellcolor[HTML]{93D4B4}0.666296 &
  \cellcolor[HTML]{9DD8BB}0.610913 \\
  \\
\\
\end{tabular}
}

{
Here, $\ast$: p-value$ < $0.05 ,$\ast \ast$: p-value$ < $0.01
,  $\ast \ast \ast$: p-value$ < $0.001 , and $\hat{citation}$: normalized citation
}
\end{table*}
\textbf{RQ2: Does the type of artifact platform influence the values of paper usage attributes? }
% (MannwhitneyuResult(statistic=9907.5, pvalue=0.11085485723423393),
 % 0.2230862831423897)

The non-parametric Mann-Whitney U test provides p-values 0.11, 0.74, and 0.56, and Cohen's D provides effect size 0.22, 0.06, and 0.03 for obtained paper usage attributes citation, normalized citation, and views/downloads of ICSE papers. 
The p-value 0.11 $>$ 0.05 fails to reject the null hypothesis that sharing artifacts on different types of artifact-sharing platforms has an equal influence on paper citation counts. 
Similarly, normalized paper citation counts and paper views/ downloads reject the null hypothesis with higher p-values. 
All of these paper usage attributes have small and negligible effect sizes. 
Thus, artifact sharing on more usage featured GitHub, or less usage featured Zenodo, Figshare, or others are not relatable to a few or more views or citations of any ICSE artifact papers.

% Normalzid (MannwhitneyuResult(statistic=9088.0, pvalue=0.7455418330414112), 0.058780398425178265) normalized
% (MannwhitneyuResult(statistic=8500.5, pvalue=0.5622248579581026),
 % 0.029869795078745428) views and downloads
\begin{table*}[htbp]
\caption{Correlation between Original and Our Approximated Artifact Views/ Downloads \& Stars}
\centering
\label{tbl_approximate_viewsdloads}
\resizebox{1.6\columnwidth}{!}{%
\begin{tabular}{llrllrl}
                         \textbf{Approximated} & \multicolumn{3}{c}{\textbf{vs Artifact views}} & \multicolumn{3}{c}{\textbf{vs Artifact downloads}} \\
\multicolumn{1}{l}{\textit{ (from GitHub)}} &
  \multicolumn{1}{c}{\textbf{p-value}} &
  \multicolumn{1}{c}{\textbf{cliff's $\delta$}} &
  \multicolumn{1}{c}{\textbf{effect-size}} &
  \multicolumn{1}{c}{\textbf{p-value}} &
  \multicolumn{1}{c}{\textbf{cliff's $\delta$}} &
  \multicolumn{1}{c}{\textbf{effect-size}} \\
has\_discussion          & *** & \cellcolor[HTML]{EB9992}-0.682331 & large  & - & \cellcolor[HTML]{F2BDB9}-0.279412 & small      \\
longest\_issue           & -   & \cellcolor[HTML]{EFAEA9}-0.451284 & medium & - & \cellcolor[HTML]{F4C5C1}-0.19339  & small      \\
\#issues\_open           & -   & \cellcolor[HTML]{FCF2F1}0.282503  & small  & - & \cellcolor[HTML]{F7D8D6}0.018343  & negligible \\
stars                    & -   & \cellcolor[HTML]{FBEFEE}0.249687  & small  & - & \cellcolor[HTML]{F9E0DE}0.105234  & negligible \\
watching                 & -   & \cellcolor[HTML]{FDF7F6}0.337563  & medium & - & \cellcolor[HTML]{FAE8E7}0.191176  & small      \\
number\_of\_commits      & -   & \cellcolor[HTML]{FEFAFA}0.373512  & medium & - & \cellcolor[HTML]{FEFAFA}0.385199  & medium     \\
longest\_pull\_request   & -   & \cellcolor[HTML]{76C8A0}0.895833  & large  & - & \cellcolor[HTML]{F1FAF5}0.483871  & large      \\
\#pull\_requests\_open   & -   & \cellcolor[HTML]{BFE5D3}0.642935  & large  & - & \cellcolor[HTML]{EEF9F4}0.492252  & large      \\
forks                    & -   & \cellcolor[HTML]{F2FAF6}0.467184  & medium & - & \cellcolor[HTML]{C5E8D7}0.63093   & large      \\
\#issues\_closed         & -   & \cellcolor[HTML]{5FBE8F}0.975564  & large  & - & \cellcolor[HTML]{A6DBC1}0.736717  & large      \\
number\_of\_contributors & -   & \cellcolor[HTML]{64C093}0.958333  & large  & - & \cellcolor[HTML]{97D5B7}0.784946  & large      \\
\#pull\_requests\_closed & -   & \cellcolor[HTML]{57BB8A}1.0         & large  & - & \cellcolor[HTML]{6BC398}0.935484  & large     \\ 
\\

\textbf{Approximated} & \multicolumn{3}{c}{\textbf{vs GitHub stars}} & \multicolumn{3}{c}{} \\
\multicolumn{1}{l}{\textit{ (from Zenodo/ Figshare)}} &
  \multicolumn{1}{c}{\textbf{p-value}} &
  \multicolumn{1}{c}{\textbf{cliff's $\delta$}} &
  \multicolumn{1}{c}{\textbf{effect-size}} &
  & &  \\
artifact-downloads       & -         & \cellcolor[HTML]{D7EFE3}0.5604690117           & large  &   & &\\
artifact-views& -& \cellcolor[HTML]{FDF7F6}0.3376884422           & medium & &  
\end{tabular}
}
\\ 
% \footnotesize
{
Here, $\ast$: p-value $< $0.05 ,$\ast \ast$: p-value$ <$ 0.01
,  $\ast \ast \ast$: p-value $< $0.001 for Mann Whitney U Test
}
\end{table*} 
\begin{table*}[htb]
\caption{ Paper vs Artifact Usages across Artifact Platforms}
\label{tbl_across_platform}
\centering
% \resizebox{\columwidth}{!}{%
\begin{tabular}{
 p{0.35\linewidth}p{0.08\linewidth} 
p{0.08\linewidth} p{0.08\linewidth}p{0.08\linewidth} 
p{0.08\linewidth} p{0.08\linewidth}  
}
 
   &
  \multicolumn{3}{c}{\textbf{p-value(Wilcoxon Signed-Rank Test)}} &
  \multicolumn{3}{c}{\textbf{effect-size(Cohen's D)}} 
 \\
  \multicolumn{1}{r}{\textbf{Paper attributes$\rightarrow$}} &
  \multicolumn{1}{p{0.05\linewidth}}{{$\hat{citation}$}} &
  \multicolumn{1}{p{0.05\linewidth}}{{citation}} &
  \multicolumn{1}{p{0.08\linewidth}}{{full views}} &
  \multicolumn{1}{p{0.05\linewidth}}{{$\hat{citation}$}} &
  \multicolumn{1}{p{0.05\linewidth}}{{citation}} &
  \multicolumn{1}{p{0.08\linewidth}}{{full views}}  \\

  \multicolumn{1}{l}{\textbf{$ \downarrow$ Artifact attributes}} &
   &&&
   &&
   \\ \hline
\#pull\_requests\_closed2artifact-downloads       & $\ast \ast \ast$ & -   & $\ast \ast \ast$ & \cellcolor[HTML]{C7E9D8}0.250386 & \cellcolor[HTML]{CCEBDC}0.214506  & \cellcolor[HTML]{EFAEA9}-0.663635 \\
\#pull\_requests\_closed2artifact-views           & $\ast \ast \ast$ & $\ast \ast \ast$ & $\ast \ast \ast$ & \cellcolor[HTML]{BBE4D0}0.32761  & \cellcolor[HTML]{BFE5D3}0.30203   & \cellcolor[HTML]{F6D3D0}-0.421118 \\
norm\_\#pull\_requests\_closed2artifact-downloads & $\ast$   & $\ast \ast \ast$ & $\ast \ast \ast$ & \cellcolor[HTML]{CAEADA}0.226675 & \cellcolor[HTML]{D9F0E5}0.128983  & -1.044821                         \\
norm\_\#pull\_requests\_closed2artifact-views     & $\ast \ast \ast$ & $\ast \ast$  & $\ast \ast \ast$ & \cellcolor[HTML]{A4DABF}0.488125 & \cellcolor[HTML]{C0E6D3}0.299659  & -1.078391                         \\
number\_of\_contributors2artifact-downloads       & $\ast \ast \ast$ & $\ast \ast \ast$ & $\ast \ast \ast$ & \cellcolor[HTML]{C6E8D7}0.257765 & \cellcolor[HTML]{CBEADB}0.221239  & \cellcolor[HTML]{EFAEA8}-0.667324 \\
number\_of\_contributors2artifact-views           & $\ast \ast \ast$ & $\ast \ast \ast$ & $\ast \ast \ast$ & \cellcolor[HTML]{86CEAB}0.686739 & \cellcolor[HTML]{8ED2B0}0.633012  & \cellcolor[HTML]{F1BAB5}-0.587105 \\
norm\_number\_of\_contributors2artifact-downloads & $\ast \ast \ast$ & -   & $\ast \ast \ast$ & \cellcolor[HTML]{C9EADA}0.233922 & \cellcolor[HTML]{D8EFE4}0.135964  & -1.042632                         \\
norm\_number\_of\_contributors2artifact-views     & $\ast \ast \ast$ & $\ast \ast \ast$ & $\ast \ast \ast$ & \cellcolor[HTML]{87CFAC}0.677568 & \cellcolor[HTML]{ABDDC4}0.440876  & -1.068855                         \\
\#issues\_closed2artifact-downloads               & $\ast \ast \ast$ & $\ast$   & $\ast \ast \ast$ & \cellcolor[HTML]{C6E8D7}0.256113 & \cellcolor[HTML]{CBEADB}0.219987  & \cellcolor[HTML]{EFAEA9}-0.662893 \\
\#issues\_closed2artifact-views                   & $\ast \ast \ast$ & $\ast \ast \ast$ & $\ast \ast \ast$ & \cellcolor[HTML]{ABDDC5}0.436185 & \cellcolor[HTML]{B1E0C9}0.398841  & \cellcolor[HTML]{F2C0BB}-0.548428 \\
norm\_\#issues\_closed2artifact-downloads         & $\ast \ast$  & $\ast \ast$  & $\ast \ast \ast$ & \cellcolor[HTML]{CAEADA}0.232399 & \cellcolor[HTML]{D8F0E4}0.134756  & -1.042534                         \\
norm\_\#issues\_closed2artifact-views             & $\ast \ast \ast$ & $\ast \ast$  & $\ast \ast \ast$ & \cellcolor[HTML]{A6DBC1}0.473248 & \cellcolor[HTML]{C0E6D3}0.29828   & -1.071274                         \\
forks2artifact-downloads                          & $\ast \ast \ast$ & $\ast \ast \ast$ & $\ast \ast \ast$ & \cellcolor[HTML]{C4E7D6}0.271763 & \cellcolor[HTML]{C7E8D8}0.252579  & \cellcolor[HTML]{F9E2E0}-0.320307 \\
forks2artifact-views                              & $\ast \ast \ast$ & $\ast \ast \ast$ & $\ast \ast \ast$ & \cellcolor[HTML]{CDEBDC}0.212086 & \cellcolor[HTML]{CDEBDC}0.207452  & \cellcolor[HTML]{E3F4EC}0.057622  \\ \hline
artifact-views2stars                              & $\ast \ast \ast$ & -   & $\ast \ast \ast$ & \cellcolor[HTML]{9AD7B9}0.549015 & \cellcolor[HTML]{DEF2E8}0.09176   & -1.152066                         \\
artifact-downloads2stars                          & $\ast \ast \ast$ & -   & $\ast \ast \ast$ & \cellcolor[HTML]{BBE4D0}0.329234 & \cellcolor[HTML]{D4EEE1}0.159439  & -1.102473                         \\
norm\_artifact-downloads2stars                    & -   & $\ast \ast \ast$ & $\ast \ast \ast$ & \cellcolor[HTML]{CEECDD}0.201218 & \cellcolor[HTML]{FCF3F2}-0.2092   & -1.182716                         \\
norm\_artifact-views2stars                        & -   & $\ast \ast \ast$ & $\ast \ast \ast$ & \cellcolor[HTML]{CFECDE}0.196871 & \cellcolor[HTML]{F5CDCA}-0.459995 & -1.196095                        \\
% \\
\end{tabular}
% \footnotesize

{
Here, $\ast$: p-value$ < $0.05 ,$\ast \ast$: p-value$ < $0.01
,  $\ast \ast \ast$: p-value $< $0.001, and $\hat{citation}$: normalized citation 
}
% }
\end{table*}
\textbf{RQ3.1: What is the correlation between artifact attributes and paper usage attributes within platforms?}

The platform-wise analyses of paper and artifact usage attributes explicitly inform us which attributes of a platform are more correlating with paper usage than other artifact usage attributes within the same platform. 
Table~\ref{tab:within_platform} shows that attributes, namely \textit{closed pull requests, longest pull requests, stars} from GitHub, and attributes namely \textit{views} and \textit{downloads} from Figshare hold the null hypotheses. 
The rest of the artifact attributes from GitHub and Figshare and all of the attributes from Zenodo show statistically significant differences against all the paper usage attributes.

Importantly, view and/or download attributes of artifacts and papers correlate with large and medium effect sizes. 
For papers' raw or normalized citations, GitHub's starring, merged pull requests, and the maximum length of artifacts' pull requests correlates with negligible or small effect sizes.

\textbf{RQ3.2:Which artifact usage attributes on a platform effectively approximate attributes of other platforms?}
\label{ans_rq2.2}

Answer to the \textbf{RQ3.1} finds only GitHub star correlates paper citations, and both artifact views and artifact downloads correlate with paper-views/downloads. 
As our experiment setup for RQ3.2 in Sec~\ref{sec_rq}, the approximation of artifact views and artifact downloads from all GitHub attributes are measured by Mann Whitney U Test and Cliff's $\delta$.
Similarly, original and approximated GitHub stars (instead of all GitHub attributes) are measured from all non-GitHub attributes.
Table~\ref{tbl_approximate_viewsdloads} shows the approximation of artifact views and downloads from the GitHub attributes, namely the numbers of \textit{pull requests closed/open, contributors, issues closed, }and\textit{ the longest pull request} are statistically similar to the original artifact views and downloads.
Notably, we can see the good approximators for artifact views/ downloads in Table~\ref{tbl_approximate_viewsdloads} intersect with GitHub attributes in Table~\ref{tab:within_platform} except for GitHub stars.
Here, the artifact-downloads are found as good approximators for GitHub stars, supporting the null hypothesis and large effect size.
This finding reflects the similarity of the selective nature of these two attributes.

\textbf{RQ3.3: Which artifact attributes, along with their approximations, exhibit correlation with paper usage attributes across platforms?}

The answers to RQ3.1 and RQ3.2 have guided us to choose the paper usage-aligned platforms-wise artifact usage attributes and their good approximator attributes for filling their missing data points.
The result of the experiment arranged for this research question is reported in Table~\ref{tbl_across_platform}.

For non-GitHub attributes, Table~\ref{tbl_across_platform} shows that only artifact downloads with their approximated values from the pull-requests closed and the normalized number of contributors show small correlations with paper citations. 
However, these are not consistent with/ without normalizations of either or both two samples.
So, artifact views or downloads with their approximated values do not show any consistent correlation with any of the paper usage attributes.
The rest of the artifact views/ downloads filled with other GitHub attributes reject the null hypothesis that artifact views/ downloads are correlated with paper citations or full views(aka. paper views/ downloads) with a p-value less than 0.001 (0.1\%) mostly.

For GitHub attributes, the stars attribute corresponds with paper citations, and the normalized GitHub stars attribute corresponds with normalized paper citations consistently.
Here, stars approximated from both artifact views and artifact downloads support the null hypothesis with small positive associations with paper citations, the impact attribute for papers.
The effect-size slightly increases when it is normalized with the paper publication year.

\textbf{Manual Analysis.}
\label{sec_result_manual}
% For our manual analysis, Table~\ref{tbl_top5_papers} shows that the mostly cited Paper~106~\citedata{DeepTest-2018-psl106} and the mostly viewed paper Paper~121 \citedata{A-2019-psl121} have appeared in the top-5 Paper~citation and Paper~view/download attributes only. 
% Next,  we explore and find that the most cited artifact, Paper~106, is a dataset artifact where most of the artifact-attributes are expected to be low and GitHub does not provide the number of views or downloads.
In our manual examination, as illustrated  
in Sec~\ref{meth_manual_analysis}, we compile a top-5 list of papers for paper and artifact usage attributes\footnote{\textbf{Our artifact is available 
 at} https://figshare.com/s/226cd9a839c1672294b3}.
 
In the top-5 lists, we observe that Paper~106~\citedata{DeepTest-2018-psl106}, which boasts the highest citation count, and Paper~121 \citedata{A-2019-psl121}, the most viewed paper, only make appearances in the top-5 lists for Paper~citation and Paper~view/download attributes respectively.
Upon further investigation, we discern that the artifact associated with the most cited paper, Paper~106~\citedata{DeepTest-2018-psl106}, is primarily a dataset paper. 
For such artifacts, most other attributes are naturally anticipated to register lower values. 
Moreover, GitHub, the platform hosting this artifact, does not offer attributes for the number of views or downloads.

We find that Paper~99~\citedata{Accurate-2018-psl99} has appeared in seven attributes' top-5 papers including the highest issues closed.
Interestingly, all of these seven are artifact-attributes. 
Similarly, Paper~145,67~\citedata{8811925,7985652} appeared for top-5 open, closed, and longest issues. Paper~1~\citedata{Building-2011-psl1} also appeared for the top-5 open and closed but did not appear in the top-5 longest issues.

The papers with the highest values for each attribute are: 
\begin{itemize}
    \item number of commits (Paper~9~\citedata{Understanding-2012-psl9}, 40162),
    \item  artifact downloads (Paper~170~\citedata{Heaps-2020-psl170}, 8388),
    \item artifact views(Paper~122~\citedata{Automated-2019-psl122}, 5463),
    \item forks (Paper~9~\citedata{Understanding-2012-psl9}, 1700),
    \item issues closed (Paper~99~\citedata{Accurate-2018-psl99}, 355),
    \item issues open (Paper~57~\citedata{Feedback-2017-psl57}, 54),
    \item longest issue (Paper~145~\citedata{8811925}, 73), 
    \item stars (Paper~135~\citedata{Guiding-2019psl135}, 365), 
    \item watching (Paper~135~\citedata{Guiding-2019psl135}, 35), 
    \item number of pull request open (Paper~143, 20),
    \item number of pull request closed (Paper~2, 472),
    \item longest pull request(Paper~148, 125),
    \item number of contributors (Paper~2,27),    
    \item paper citations (Paper~106~\citedata{DeepTest-2018-psl106}, 321),  and
    \item paper views (Paper~121~\citedata{A-2019-psl121}, 5704).

\end{itemize}
\begin{figure}[htb]
    \centering
    \includegraphics[width=.95\linewidth]{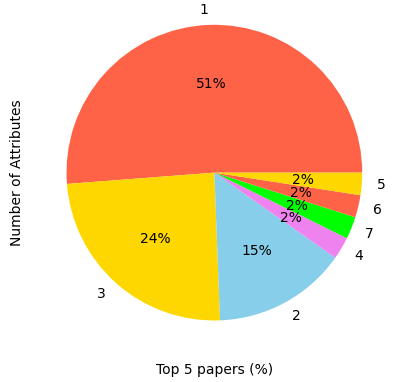}
    \caption{Top-5 Papers and Paper/Artifact Attributes}
    \label{fig_pie}
\end{figure}
% The other papers that have appeared in top-5 of only one attribute are :
% % Paper~
% % 3, 5, 11,12,15, 16,17,18,19,20, 21, 31, 32; 
% Paper 148~\citedata{The-2019-psl148}, Paper 49~\citedata{Code-2017-psl49}, Paper 32~\citedata{MU-MNINT-2015-psl32}, Paper 37~\citedata{DoubleTake-2016-psl37}, Paper 58~\citedata{How-2017-psl58}, Paper 218~\citedata{Interface-2021-psl218}, Paper 170~\citedata{Heaps-2020-psl170}, Paper 233~\citedata{SOAR:-2021-psl233}, Paper 228~\citedata{Representation-2021-psl228}, Paper 230~\citedata{Semi-supervised-2021-psl230}, Paper 93~\citedata{Measuring-2018-psl93}, Paper 131~\citedata{Gigahorse:-2019-psl131}, Paper 122~\citedata{Automated-2019-psl122}, Paper 149~\citedata{The-2019-psl149}, Paper 102~\citedata{A-2018-psl102}
% ;
% in top-5 of two attributes are:
% % 22,23, Paper 29, Paper 30.   
%  Paper 212~\citedata{FlakeFlagger:-2021-psl212}, Paper 17~\citedata{Data-2013-psl17}, Paper 121~\citedata{A-2019-psl121}, Paper 70~\citedata{Semantically-2017-psll70}, Paper 9~\citedata{Understanding-2012-psl9}, Paper 34~\citedata{On-2015-psl34}, Paper 106~\citedata{DeepTest-2018-psl106}, Paper 45~\citedata{Termination-2016-psl45}, Paper 57~\citedata{Feedback-2017-psl57}.

% Table~\ref{tbl_top5_papers} also highlights the papers that have the highest values of various attribute values from both paper and artifact attributes.

Fig.~\ref{fig_pie} shows the overall percentage of ICSE papers that appeared in the top-5 paper/artifacts attribute values.
Half of them have come from a single attribute's top-5 list covering all platform types, including the paper, GitHub, and non-GitHub.
This shows the diversity of the usage of these attributes.

We also see that 39\% of papers from the top 5 lists co-appear in attributes from the same platforms covering all platforms.
Importantly, even though we see papers co-appear at the top list from GitHub and paper usage attributes, top papers from paper usage do not co-appear in any non-GitHub (i.e., Zenodo or Figshare) attributes.
One likely cause is the number of attributes from GitHub is greater than the number of attributes from others.

% We see papers that have appeared in attributes from both paper and artifact usages having paper citation in common are: Paper~82~\citedata{Deep-2018-psl82}(with longest issue, stars) and  Paper~135~\citedata{Guiding-2019psl135}(with forks, highest stars and highest watching). 
% This finding corroborates the finding in above statistical analysis in Sec~\ref{sec_res_stat_analysis}.
Interestingly, papers that consistently emerge across both paper and artifact usage attributes, with paper citations as a shared attribute, include Paper~82~\citedata{Deep-2018-psl82} (highlighted for the longest issue and number of stars) and Paper~135~\citedata{Guiding-2019psl135} (noted for its number of forks, highest stars, and highest watch count). 
This observation aligns with and reinforces the insights from the statistical analysis.% detailed in Sec~\ref{sec_res_stat_analysis}.

% \input{res/tbl/tbl_metric_pvalue}

% \input{res/tbl/tbl_metrics}

% \section{Related Work}
% \label{sec_related}
% \input{sec_relatedwork}

\section{Threats to Validity}
\label{sec_threats}
In this section, we identify potential threats to the validity of our research and elaborate on the strategies we employed in our research design to address these challenges.

% \subsection{External Validity} 
% The external validity examines the generalizability of findings.
% The major threat to generalizability is our data collection. 
% To address this issue, we consider the top and the most prominent software conference venue ICSE.
% This can be generalizable to other conferences that carry out similar provisions for artifact badges.
\textbf{External validity }concerns the broad applicability of our findings. 
The primary challenge to ensuring this validity arises from our data collection method. 
Inspired by the literature,
we focus on ICSE, a leading and highly regarded software conference venue to mitigate this. 
Our findings can likely be extrapolated to other conferences that adopt similar practices for awarding artifact badges.
% However, the findings can bring different results for different conference papers.  

% \subsection{Internal Validity}
% Internal validity examines the treatment and outcome of a study. 
% In our study, we adopt a suitable statistical significance testing, two samples paired \textit{t}-tests, 
% for statistical analysis. 
% The non-rejection to the null hypothesis does not necessarily mean the resemblance of the pair of attribute values. 
% However, our further manual analysis supports the resemblance. 
% So, this mitigates the threat to our outcomes.
% Another potential threat in our study is usage-count. 
% We overcome this threat by normalizing their values considering their age which is the time-span between their publication year and the current year. 
\textbf{Internal validity} assesses the relationship between a study's treatment and its results. 
We have employed an appropriate statistical method for our analysis: the non-parametric \textit{t}-tests. 
Notably, failing to reject the null hypothesis does not inherently confirm the similarity between the paired attribute values. 
Nevertheless, our supplementary manual analysis bolsters the argument for similarity, thereby reducing potential concerns about our conclusions. 
Another conceivable challenge in our study relates to usage count. 
To counteract this, we have normalized these counts, taking into account the duration since their publication up to 2023.%the present year.

% \subsection{Construct Validity}
% Construct validity concerns with theory and observations. 
% In our study, we find the artifact-usage attribute values are missing as they come from different platforms that are featured differently. 
% To tackle this threat, we opt for an entirely customized missing data approximation instead of popular and available techniques.
\textbf{Construct validity} deals with the alignment between theoretical constructs and empirical observations. 
In our research, we noticed that some artifact-usage attribute values were absent, primarily due to the diverse features of different platforms from which the data was sourced. 
We chose a tailored approach to approximate missing data to address this potential issue rather than relying on widely used and established techniques.

\section{ Conclusion}
\label{sec_conclusion}
While ICSE recognizes and honors papers, artifacts, and individuals annually, it distinctively acknowledges the ``ICSE N-10 Most Influential Paper". 
Yet, there exists no parallel accolade for ``ICSE N-10 Influential Artifacts". 
It is essential to clarify that our research does not aim to pinpoint ICSE papers whose associated artifacts have significantly impacted software engineering over a decade. 
Instead, our study aims to identify promising attributes associated with artifact usage that could provide insights into the influence these artifacts exert. 
We undertake this insight by comparing all available artifact usage attributes with a universally accepted quantitative gauge of paper impact: paper citations. 
Throughout this exploration, the attributes indicating artifact usage diverge across different platforms. 

Our combined approach of statistical and manual analyses reveals that the artifact attributes closely mirroring paper citations originate from a range of artifact-sharing platforms. 
This observation underscores the need for either innovative platforms or modifications to existing ones to evaluate the influence of research artifacts proficiently.
Our paper marks a pioneering effort to highlight the importance of artifacts in top-tier research, aspiring to galvanize and benefit the global software engineering fraternity.

\bibliographystyle{IEEEtran}
\bibliography{references}

\begin{thebibliography}{10}
\providecommand{\url}[1]{#1}
\csname url@samestyle\endcsname
\providecommand{\newblock}{\relax}
\providecommand{\bibinfo}[2]{#2}
\providecommand{\BIBentrySTDinterwordspacing}{\spaceskip=0pt\relax}
\providecommand{\BIBentryALTinterwordstretchfactor}{4}
\providecommand{\BIBentryALTinterwordspacing}{\spaceskip=\fontdimen2\font plus
\BIBentryALTinterwordstretchfactor\fontdimen3\font minus
  \fontdimen4\font\relax}
\providecommand{\BIBforeignlanguage}[2]{{%
\expandafter\ifx\csname l@#1\endcsname\relax
\typeout{** WARNING: IEEEtran.bst: No hyphenation pattern has been}%
\typeout{** loaded for the language `#1'. Using the pattern for}%
\typeout{** the default language instead.}%
\else
\language=\csname l@#1\endcsname
\fi
#2}}
\providecommand{\BIBdecl}{\relax}
\BIBdecl

\bibitem{liem2023treat}
C.~C.~S. Liem and A.~M. Demetriou, ``Treat societally impactful scientific
  insights as open-source software artifacts,'' in \emph{2023 IEEE/ACM 45th
  International Conference on Software Engineering: Software Engineering in
  Society (ICSE-SEIS)}, 2023, pp. 150--156.

\bibitem{Mendez_Graziotin_Wagner_Seibold_2020openscienceinse}
D.~Mendez, D.~Graziotin, S.~Wagner, and H.~Seibold,
  \emph{\BIBforeignlanguage{en}{Open Science in Software Engineering}}.\hskip
  1em plus 0.5em minus 0.4em\relax Cham: Springer International Publishing,
  2020, p. 477–501.

\bibitem{heumuller2020publish}
R.~Heum{\"u}ller, S.~Nielebock, J.~Kr{\"u}ger, and F.~Ortmeier, ``Publish or
  perish, but do not forget your software artifacts,'' \emph{Empirical Software
  Engineering}, vol.~25, no.~6, pp. 4585--4616, 2020.

\bibitem{frachtenberg2022research}
E.~Frachtenberg, ``Research artifacts and citations in computer systems
  papers,'' \emph{PeerJ Computer Science}, vol.~8, p. e887, 2022.

\bibitem{stefan2022retrospective}
S.~Winter, C.~S. Timperley, B.~Hermann, J.~Cito, J.~Bell, M.~Hilton, and
  D.~Beyer, ``A retrospective study of one decade of artifact evaluations,'' in
  \emph{Proceedings of the 30th ACM Joint European Software Engineering
  Conference and Symposium on the Foundations of Software Engineering}, ser.
  ESEC/FSE 2022.\hskip 1em plus 0.5em minus 0.4em\relax New York, NY, USA: ACM,
  2022.

\bibitem{ahmed2022automatic}
S.~Ahmed, A.~Ahmed, and N.~U. Eisty, ``Automatic transformation of natural to
  unified modeling language: A systematic review,'' in \emph{IEEE/ACIS 20th
  International Conference on Software Engineering Research, Management \&
  Applications}.\hskip 1em plus 0.5em minus 0.4em\relax IEEE, 2022, pp.
  112--119.

\bibitem{mann1947test}
H.~B. Mann and D.~R. Whitney, ``On a test of whether one of two random
  variables is stochastically larger than the other,'' \emph{The annals of
  mathematical statistics}, pp. 50--60, 1947.

\bibitem{wilcoxon1970critical}
F.~Wilcoxon, S.~Katti \emph{et~al.}, ``Critical values and probability levels
  for the wilcoxon rank sum test and the wilcoxon signed rank test,''
  \emph{Selected tables in mathematical statistics}, vol.~1, pp. 171--259,
  1970.

\bibitem{2020SciPy-NMeth}
P.~Virtanen, et~al, and {SciPy 1.0 Contributors}, ``{{SciPy} 1.0: Fundamental
  Algorithms for Scientific Computing in Python},'' \emph{Nature Methods},
  vol.~17, 2020.

\bibitem{cliff1993dominance}
N.~Cliff, ``Dominance statistics: Ordinal analyses to answer ordinal
  questions.'' \emph{Psychological bulletin}, vol. 114, no.~3, p. 494, 1993.

\bibitem{cohen2013statistical}
J.~Cohen, \emph{Statistical power analysis for the behavioral sciences}.\hskip
  1em plus 0.5em minus 0.4em\relax Academic press, 2013.

\bibitem{hess2004robust}
M.~R. Hess and J.~D. Kromrey, ``Robust confidence intervals for effect sizes: A
  comparative study of cohen’sd and cliff’s delta under non-normality and
  heterogeneous variances,'' in \emph{annual meeting of the American
  Educational Research Association}, vol.~1.\hskip 1em plus 0.5em minus
  0.4em\relax Citeseer, 2004.

\end{thebibliography}
\end{document}